\documentclass[prb,twocolumn,showpacs,preprintnumbers,amsmath,amssymb,superscriptaddress]{revtex4}
\usepackage{graphicx}
\usepackage{amssymb}
\usepackage{natbib}
\begin{document}
\title{Magnetic anisotropy in hole-doped superconducting Ba$_{0.67}$K$_{0.33}$Fe$_2$As$_{2}$ probed by polarized inelastic neutron scattering}

\author{Chenglin Zhang}
\affiliation{Department of Physics and Astronomy, The University of
Tennessee, Knoxville, Tennessee 37996-1200, USA}

\author{Mengshu Liu}
\affiliation{Department of Physics and Astronomy, The University of
Tennessee, Knoxville, Tennessee 37996-1200, USA}

\author{Yixi Su}
 \affiliation{
J\"{u}lich Centre for Neutron Science JCNS-FRM II, Forschungszentrum J\"{u}lich GmbH, Outstation at FRM II, Lichtenbergstrasse 1, D-85747 Garching, Germany 
}

\author{Louis-Pierre Regnault}
\affiliation{Institut Laue-Langevin, 6, rue Jules Horowitz, BP 156,
38042 Grenoble Cedex 9, France}

\author{Meng Wang}
\affiliation{Department of Physics and Astronomy, The University of
Tennessee, Knoxville, Tennessee 37996-1200, USA}
\affiliation{Beijing National Laboratory for Condensed Matter Physics,
Institute of Physics, Chinese Academy of Sciences, Beijing 100190, China}

\author{Guotai Tan}
\affiliation{Department of Physics and Astronomy, The University of
Tennessee, Knoxville, Tennessee 37996-1200, USA}
\affiliation{Physics Department, Beijing Normal University, Beijing 100875, China}

\author{Th. Br$\rm \ddot{u}$ckel}
\affiliation{J$\ddot{u}$lich Centre for Neutron Science JCNS and Peter Gr$\ddot{u}$nberg Institut PGI, JARA-FIT, Forschungszentrum J$\ddot{u}$lich GmbH, 52425 J$\ddot{u}$lich, Germany}

\author{Takeshi Egami}
 \affiliation{Department of Physics and Astronomy, The University of Tennessee,
Knoxville, Tennessee 37996-1200, USA}
\affiliation{Department of Materials Science and Engineering, The University of Tennessee,
Knoxville, Tennessee 37996-1200, USA}
\affiliation{Joint Institute of Neutron Sciences, Oak Ridge National Laboratory,
Oak Ridge, Tennessee 37831, USA}

\author{Pengcheng Dai}
\email{pdai@utk.edu }
 \affiliation{Department of Physics and Astronomy, The University of Tennessee,
Knoxville, Tennessee 37996-1200, USA}
\affiliation{Beijing National Laboratory for Condensed Matter Physics,
Institute of Physics, Chinese Academy of Sciences, Beijing 100190, China}

\date{\today}
\begin{abstract}
We use polarized inelastic neutron scattering (INS) to study spin excitations of optimally 
hole-doped superconductor Ba$_{0.67}$K$_{0.33}$Fe$_2$As$_{2}$ ($T_c=38$ K).
 In the normal
state, the imaginary part of the dynamic susceptibility,
$\chi^{\prime\prime}(Q,\omega)$, shows magnetic anisotropy for energies below $\sim$7 meV
with $c$-axis polarized spin excitations 
larger than that of the in-plane component.  Upon entering into the superconducting state,
previous unpolarized INS experiments have shown that spin gaps 
at $\sim$5 and 0.75 meV open at wave vectors $Q=(0.5,0.5,0)$ and $(0.5,0.5,1)$, respectively, with 
a broad neutron spin resonance at $E_r=15$ meV.  
Our neutron polarization analysis reveals that the large difference in spin gaps is purely 
due to different spin gaps in the $c$-axis and in-plane polarized spin excitations, resulting 
resonance with different energy widths for the $c$-axis and in-plane spin excitations.  
The observation of spin anisotropy in both opitmally electron and hole-doped BaFe$_2$As$_2$ is due to 
their proximity to the AF ordered BaFe$_2$As$_2$ where
spin anisotropy exists below $T_N$.
\end{abstract}
\maketitle

Neutron polarization analysis has played an important role in determining the magnetic structure and excitations of solids \cite{moon}.  For high-transition temperature (High-$T_c$) copper oxide superconductors derived from hole or electron-doping from their antiferromagnetic (AF) parent compounds, neutron polarization analysis have conclusively shown that the collective magnetic excitation coupled to superconductivity 
at the AF wave vector of the parent compounds,
termed neutron spin resonance \cite{mignod}, has a magnetic origin \cite{mook,fong,dai96,headings,wilson,jzhao11,eschrig}. Furthermore, by carrying out neutron polarization analysis with
a spin-polarized incident neutron beam along the scattering wave vector ${\bf Q}= {\bf k}_i-{\bf k}_f$ (where ${\bf k}_i$ and ${\bf k}_f$ are the incident and final wave vectors of the neutron, respectively), $\hat{{\bf x}}||{\bf Q}$; perpendicular to ${\bf Q}$ but in the scattering plane, $\hat{{\bf y}}\bot{\bf Q}$; and
perpendicular to ${\bf Q}$ and the scattering plane, $\hat{{\bf z}}\bot{\bf Q}$, one can 
use neutron spin flip (SF) scattering cross sections $\sigma_{xx}^{\rm SF},\sigma_{yy}^{\rm SF},$ and $\sigma_{zz}^{\rm SF}$ to 
determine the spatial anisotropy of spin excitations \cite{moon}. If the resonance is an
isotropic triplet excitation of the singlet superconducting ground
state, one expects that the degenerate triplet would be isotropic
in space as pure paramagnetic scattering \cite{eschrig}.  For optimally hole-doped copper oxide superconductor
YBa$_2$Cu$_3$O$_{6.9}$ ($T_c=93$ K), neutron polarization analysis reveals that spin excitations
in the normal state are spatially isotropic and 
featureless for energies $10\leq E\leq 60$ meV, consistent with pure paramagnetic scattering.  Upon entering into the superconducting state, a quasi-isotropic spin resonance occurs at $E_r=40$ meV to within the precision
of the measurements
and a spin anisotropy develops in the lower energy $10\leq E\leq 30$ meV, resulting in a clear spin gap below 22 meV 
for the $c$-axis polarized dynamic susceptibility $\chi^{\prime\prime}_c$ and 
in-plane $\chi^{\prime\prime}_{a/b}$ for $E\geq 10$ meV \cite{headings}.  
The low-energy spin anisotropy is likely due to spin-orbit coupling in the system.
For optimally electron-doped copper 
oxide superconductor Pr$_{0.88}$LaCe$_{0.12}$CuO$_{4-\delta}$, spin excitations are isotropic both above and below $T_c$ \cite{jzhao11}.  Therefore, the spin anisotropy in the superconducting state of hole-doped  
YBa$_2$Cu$_3$O$_{6.9}$ is unrelated to the normal state paramagnetic scattering.

Like copper oxide superconductors, superconductivity in iron pnictides also
arises when electrons or holes are doped into their AF parent 
compounds \cite{Kamihara,Rotter,ljli,Cruz,dai}. Furthermore, unpolarized neutron scattering
experiments have shown that both hole and electron-doped iron pnictides exhibits a neutron spin
resonance similar to copper oxide superconductors \cite{Christianson,clzhang,lumsden,chi,inosov,hqluo12}.
In the initial
polarized neutron scattering experiment on 
optimally electron-doped superconductor
BaFe$_{1.9}$Ni$_{0.1}$As$_2$ ($T_c=20$ K),
$\chi^{\prime\prime}_c$ was found to be much larger than $\chi^{\prime\prime}_{a/b}$
for energies $2\leq E\leq 6$ meV below $T_c$ , while the resonance at $E_r=7$ meV
is only weakly anisotropic \cite{lipscombe10}. In a subsequent polarized neutron 
scattering measurement on undoped AF parent 
compound BaFe$_2$As$_2$ \cite{qureshi12}, isotropic paramagnetic scattering 
at low-energy ($E=10$ meV) were found to
become anisotropic spin waves below the N${\rm \acute{e}}$el temperature $T_N$ with 
a much larger in-plane ($\chi^{\prime\prime}_{a/b}$) spin gap 
than that of the out-of-plane gap ($\chi^{\prime\prime}_c$).
These results indicate a strong single-ion anisotropy and spin-orbit coupling, suggesting that
more energy is needed to rotate a spin within the orthorhombic $a$-$b$ plane than rotating it
to the $c$-axis \cite{qureshi12}.  However, similar polarized neutron experiments on electron-overdoped 
BaFe$_{1.85}$Ni$_{0.15}$As$_2$ ($T_c=14$ K), which is far away from the AF ordered phase, 
reveal isotropic paramagnetic scattering both above and below $T_c$ \cite{msliu12}. Very recently, Steffens {\it et al.} report evidence for two resonance-like excitations in the superconducting state of optimally electron-doped BaFe$_{1.88}$Co$_{0.12}$As$_2$ ($T_c=24$ K).  In addition to an isotropic resonance at $E=8$ meV
with weak dispersion along the $c$-axis, there is a resonance at $E=4$ meV polarized only along the $c$-axis 
with strong intensity variation along the $c$-axis \cite{steffens}.
In the normal state, there are isotropic paramagnetic scattering at AF wave vectors with $L=0$ and weak anisotropic scattering with a larger $c$-axis polarized 
intensity at $L=1$ \cite{steffens}.  

\begin{figure}
\includegraphics[scale=.45]{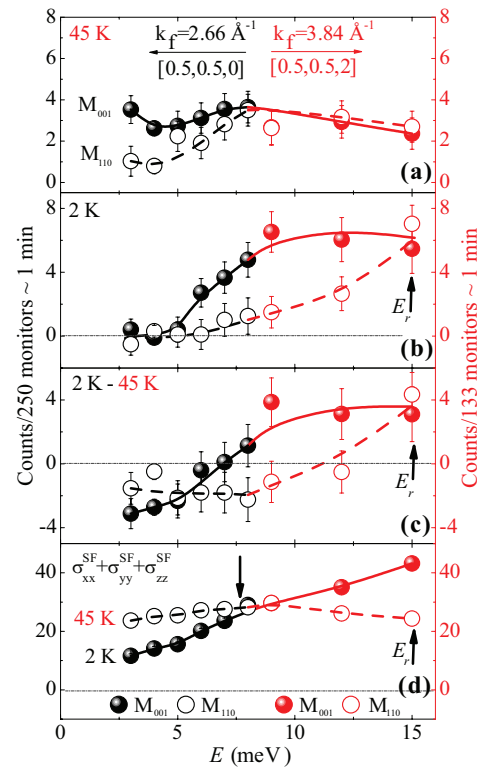}
\caption { (Color online)  Neutron polarization analysis determined 
$c$-axis ($\chi^{\prime\prime}_c\propto M_{001}$) and in-plane ($\chi^{\prime\prime}_{a/b}\propto M_{110}$)
components of spin excitations in Ba$_{0.67}$K$_{0.33}$Fe$_2$As$_{2}$ from raw SF constant-$Q$ scans at
${\bf Q}=(0.5,0.5,0)$ and and $(0.5,0.5,2)$.  
To extract $M_{001}$ and $M_{110}$, we use methods described in Ref. \cite{msliu12} and
assume $M_{1\bar{1}0}=M_{110}$ in the tetragonal crystal.
(a) Energy dependence of $M_{001}$ and $M_{110}$  
at $T=45$ K. (b) Identical scans at $T=2$ K. (c) The solid and
open circles show the temperature difference (2 K$-$45 K) for 
$M_{001}$ and $M_{110}$, respectively.  (d) The sum 
of $\sigma_{xx}^{\rm SF}+\sigma_{yy}^{\rm SF}+\sigma_{zz}^{\rm SF}$ at 45 and 2 K.  
Since background scattering is not expected to change between these temperatures \cite{clzhang}, 
such a procedure will increase statistics of 
magnetic scattering.
The black data points are
collected at ${\bf Q}=(0.5,0.5,0)$ with $k_f=2.66$ \AA$^{-1}$, while the red data points
are at ${\bf Q}=(0.5,0.5,1)$ with $k_f=3.84$ \AA$^{-1}$. 
The solid and dashed lines are guided to the eyes.
 }
\end{figure}

\begin{figure}
\includegraphics[scale=.45]{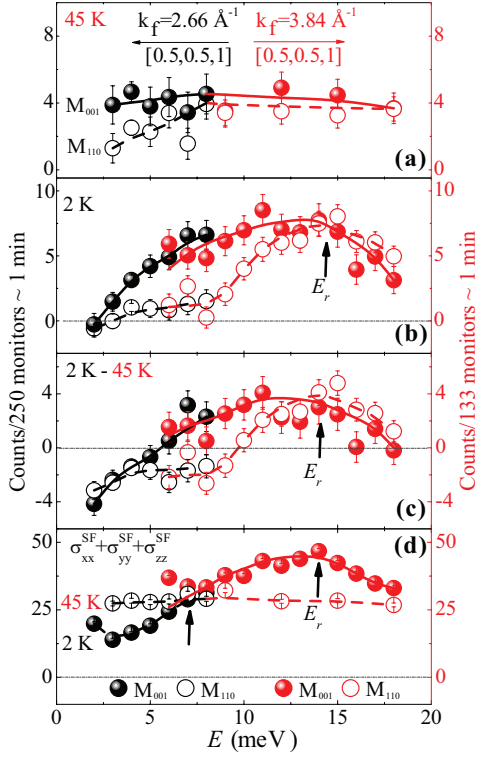}
\caption { (Color online)  Constant-$Q$ scans at ${\bf
Q}=(0.5,0.5,1)$ below and above $T_c$. 
(a) Energy dependence of $M_{001}$ and $M_{110}$
at $T=45$ K and (b) at 2 K.  The superconductivity-induced
spin gaps are at $\leq 2$ and 7 meV for $M_{001}$ and $M_{110}$, respectively.  At
resonance energy of $E_r=15$ meV, the scattering is isotropic.
(c) The solid and
open circles show the temperature difference (2 K$-$45 K) for 
$M_{001}$ and $M_{110}$, respectively.  (d) The sum 
of $\sigma_{xx}^{\rm SF}+\sigma_{yy}^{\rm SF}+\sigma_{zz}^{\rm SF}$ at 45 and 2 K. The solid and dashed lines are guided to the eyes.
 }
\end{figure}

If the observed anisotropic magnetic scattering in the superconducting state of optimally electron-doped 
BaFe$_{1.9}$Ni$_{0.1}$As$_2$ \cite{lipscombe10} and BaFe$_{1.88}$Co$_{0.12}$As$_2$ \cite{steffens}
are indeed associated with the anisotropic spin waves in BaFe$_2$As$_2$ \cite{qureshi12}, 
one would expect similar anisotropic spin excitations in hole-doped materials not too far away from the parent compound.
In this paper, we report neutron polarization analysis on spin excitations of 
the optimally hole-doped superconducting
Ba$_{0.67}$K$_{0.33}$Fe$_2$As$_{2}$.
From the previous unpolarized INS work on the same sample, 
we know that spin excitations in
the superconducting state have a resonance at $E_r=15$ meV, a small spin gap ($E_g\approx 0.75$ meV) at 
${\bf Q}=(0.5,0.5,0)$ and a large gap ($E_g=5$ meV) at $(0.5,0.5,1)$ \cite{clzhang}. 
In the normal state, spin excitations at both wave vectors are gapless and increase linearly with increasing energy \cite{clzhang}.
Our polarized INS experiments reveal that the persistent low-energy spin excitations at 
the AF wave vector $(0.5,0.5,1)$ below $T_c$ are entirely $c$-axis polarized.  
Although there is also superconductivity-induced spin anisotropy similar to optimally 
electron-doped  BaFe$_{1.9}$Ni$_{0.1}$As$_2$ \cite{lipscombe10} and BaFe$_{1.88}$Co$_{0.12}$As$_2$ \cite{steffens}, the low-energy $c$-axis polarized spin excitations do not change across $T_c$ and therefore
cannot have the same microscopic origin as the spin isotropic resonance at $E_r=15$ meV. We 
suggest that the persistent $c$-axis polarized spin excitations 
in the superconducting state of optimally hole and electron-doped iron pnictide superconductors 
is due to their proximity to the AF ordered parent compound. Their coupling to superconductivity may 
arise from different contributions of 
Fe 3$d_{X^2-Y^2}$ and 3$d_{XZ/YZ}$ orbitals to superconductivity \cite{malaeb}.

\begin{figure}
\includegraphics[scale=.45]{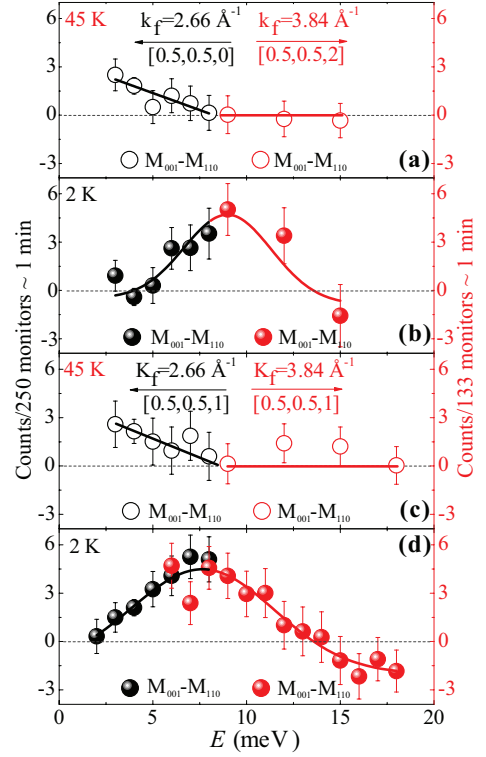}
\caption { (Color online) 
Energy dependence of spin anisotropy as determined by 
the difference between $M_{001}-M_{110}$ for temperatures (a) 45 K
and (b) 2 K at wave vector ${\bf Q}=(0.5,0.5,0)$ and ${\bf Q}=(0.5,0.5,2)$ .
Similar differences above (c) and below (d) $T_c$ at 
${\bf Q}=(0.5,0.5,1)$.  The energy width is broader in (d) 
compared with (b). The solid and dashed lines are guided to the eyes.
 }
\end{figure}

Single crystals of Ba$_{0.67}$K$_{0.33}$Fe$_2$As$_{2}$ are grown by a self-flux
method \cite{clzhang}. About 10 grams of single crystals are coaligned in the $[H,H,L]$ scattering plane
(with mosaicity $3^\circ$ at full width half maximum) with a
tetragonal unit cell for which $a=b=3.93$ \AA, and $c=13.29$ \AA.
In this notation, the vector \textbf{Q} in three-dimensional
reciprocal space in \AA$^{-1}$ is defined as
$\textbf{Q}=H\textbf{a} ^*+K\textbf{b} ^*+L\textbf{c} ^*$, where
$H$, $K$, and $L$ are Miller indices and $\textbf{a}
^*=\hat{\textbf{a}}2\pi/a, \textbf{b}
^*=\hat{\textbf{b}}2\pi/b,\textbf{c} ^*=\hat{\textbf{c}}2\pi/c$
are reciprocal lattice vectors. 
Our polarized INS experiments were carried out 
on the IN22 triple-axis spectrometer with Cryopad capability at the Institut Laue-Langevin
in Grenoble, France.
The fixed final neutron wave vectors were set at $k_f = 2.66$ \AA$^{-1}$ and 
$k_f = 3.84$ \AA$^{-1}$ in order to close the scattering triangles. 
To compare with previous polarized INS 
results on iron pnictides \cite{lipscombe10,qureshi12,msliu12,steffens}, 
we converted the measured neutron SF scattering cross sections 
$\sigma_{xx}^{\rm SF}$, $\sigma_{yy}^{\rm SF}$, and $\sigma_{zz}^{\rm SF}$ 
into $c$-axis ($M_{001}$) 
and in-plane ($M_{110}$) components of the 
magnetic scattering \cite{msliu12}.

\begin{figure}
\includegraphics[scale=.35]{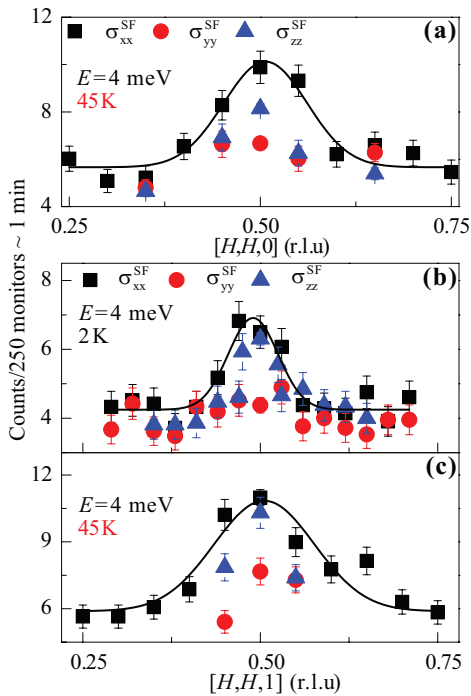}
\caption { (Color online)
Constant-energy scans along the $[H,H,0]$ and $[H,H,1]$ directions at
an energy transfer of $E=4$ meV for different neutron polarization directions.
(a) Neutron SF scattering cross sections $\sigma_{xx}^{\rm SF}$, $\sigma_{yy}^{\rm SF}$,
and $\sigma_{zz}^{\rm SF}$ at 45 K along the $[H,H,0]$ direction. 
Similar scans along the $[H,H,1]$ direction at (b) 2 K and (c) 45 K.  All data are obtained with
$k_f=2.66\ {\rm \AA}^{-1}$. The solid lines are fit by Gaussian.
}
\end{figure}

Figure 1 shows energy scans above and below $T_c$ at wave vectors ${\bf Q}=(0.5,0.5,0)$
and $(0.5,0.5,2)$.  We chose these two equivalent wave vectors with different fixed final neutron 
energies to satisfy the kinematic condition for the large covered energy range.
Since the iron magnetic form factors, geometrical factors, and instrumental resolutions are 
different at these two wave vectors, we use left and right scales for 
${\bf Q}=(0.5,0.5,0)$ and $(0.5,0.5,2)$, respectively.  In the normal state
(45 K), spin anisotropy for energies below $E\approx 7$ meV is clear
with $M_{001}$ ($\chi^{\prime\prime}_c$) larger than $M_{110}$ ($\chi^{\prime\prime}_{a/b}$) [Fig. 1(a)].
For $E > 7$ meV, spin excitations are nearly isotropic.  This is different from electron-doped 
BaFe$_{1.88}$Co$_{0.12}$As$_2$, where paramagnetic scattering at ${\bf Q}=(0.5,0.5,0)$
 is isotropic above $T_c$ \cite{steffens}. In the superconducting state (2 K),
$M_{001}$ and $M_{110}$ in Ba$_{0.67}$K$_{0.33}$Fe$_2$As$_{2}$
vanish below 5 meV, consistent with opening of 
a superconductivity-induced spin gap [Fig. 1(b)] \cite{clzhang}.  From $E=5$ meV to the 
resonance energy at $E_r=15$ meV, both $M_{001}$ and $M_{110}$ increase with increasing energy, but with 
different slope resulting significant spin 
anisotropy ($M_{001}>M_{110}$) appearing near $E\approx 8$ meV [Fig. 1(b)].  This
is similar to the spin anisotropy in BaFe$_{1.88}$Co$_{0.12}$As$_2$ \cite{steffens}. Figure 1(c) shows
the temperature difference of magnetic scattering, revealing net intensity gains
for $M_{001}$ and $M_{110}$ only above $\sim$7 and 10 meV, respectively.
Figure 1(d) shows the sum of the SF magnetic scattering 
intensities for three different neutron polarizations, which improve 
the statistics, above and below $T_c$.  Consistent with Fig. 1(c), the superconductivity-induced
net magnetic intensity gain appears only above $\sim$7 meV, forming a resonance at $E_r=15$ meV.

Figure 2 summarizes the identical scans as that of Fig. 1 at the AF wave vector
${\bf Q}=(0.5,0.5,1)$ above and below $T_c$.  
At $T=45$ K, we see clear spin anisotropy below $E\approx 7$ meV with $M_{001}>M_{110}$ similar to
the spin excitations at 
${\bf Q}=(0.5,0.5,0)$ [Fig. 2(a)].  Upon cooling to 2 K, a large spin gap opens below $E\approx 7$ meV in
$M_{110}$, but there is still magnetic scattering in $M_{001}$ extending 
to at least $2$ meV.  Therefore, the low-energy signal above $\sim$1 meV at ${\bf Q}=(0.5,0.5,1)$
reported in the earlier unpolarized neutron 
measurements \cite{clzhang} are entirely $c$-axis polarized magnetic scattering.
The neutron spin resonance at $E_r=15$ is isotropic. The temperature
difference plots between 2 and 45 K display a broad 
and narrow peak for $M_{001}$ and $M_{110}$, respectively [Fig. 2(c)].
Fig. 2(d) shows the sum of SF magnetic scattering below and above $T_c$.  Consistent with
unpolarized work \cite{clzhang}, we see net intensity gain of the resonance 
in the superconducting state for energies above $E\approx 7$ meV, different from
that of BaFe$_{1.88}$Co$_{0.12}$As$_2$ where the magnetic intensity starts to gain
 above $E=4$ meV in the superconducting state [Fig. 4(b) in \cite{steffens}].

To further illustrate the effect of spin anisotropy, we plot in Figs. 3(a)-3(d) the
differences of $(M_{001}-M_{110})$ above and below $T_c$ at wave vectors 
${\bf Q}=(0.5,0.5,0)$ and $(0.5,0.5,1)$.  In the normal state, we see 
clear magnetic anisotropy with $M_{001}>M_{110}$ for energies below $\sim$7 meV
[Figs. 3(a) and 3(c)]. In the superconducting state, the 
$(M_{001}-M_{110})$ differences reveal similar intensity peaks centered around  
$\sim$7 meV at ${\bf Q}=(0.5,0.5,0)$ and $(0.5,0.5,1)$, but with a much broader width for
${\bf Q}=(0.5,0.5,1)$ [Figs. 3(b) and 3(d)].  Since there are essentially no 
intensity gain in $M_{001}$ across $T_c$ near $\sim$7 meV [Figs. 1(c) and 2(c)], 
the apparent peaks in $(M_{001}-M_{110})$ arise from different responses of  
$M_{001}$ and $M_{110}$ across $T_c$.  While the intensity of $M_{001}$ across $T_c$ 
is suppressed below $\sim$7 meV and enhanced above it, similar cross over energy occurs 
 around 10 meV in $M_{110}$, thus resulting peaks near 7 meV in $(M_{001}-M_{110})$ at 2 K [Figs. 3(b) and 3(d)].
Therefore, the differences in superconductivity-induced spin gaps in 
$M_{001}$ and $M_{110}$ at ${\bf Q}=(0.5,0.5,0)$ and $(0.5,0.5,1)$ are 
causing peaks in $(M_{001}-M_{110})$.

Finally, to confirm the low-energy spin anisotropy 
discussed in Figs. 1-3, we show in Figs. 4(a)-4(c) constant-energy scans with three different neutron 
polarizations at $E=4$ meV along the $[H,H,0]$ and
$[H,H,1]$ directions.  In the normal state, $\sigma_{xx}^{\rm SF}$ shows clear peaks
at ${\bf Q}=(0.5,0.5,0)$ and $(0.5,0.5,1)$ [Figs. 4(a) and 4(c)].  
In both cases, we also find
$\sigma_{xx}^{\rm SF}\ge \sigma_{zz}^{\rm SF}>\sigma_{yy}^{\rm SF}$, thus confirming the
anisotropic nature
of the magnetic scattering with $M_{001}>M_{110}$. In the superconducting state, 
while $\sigma_{xx}^{\rm SF}$ and $\sigma_{zz}^{\rm SF}$ are peaked at $(0.5,0.5,1)$,  
$\sigma_{yy}^{\rm SF}$ is featureless.  These results again 
confirm the presence of a larger superconductivity-induced 
spin gap in $M_{110}$ than that in $M_{001}$ [Fig. 2(b)].

From Figs. 1-4, we see anisotropic spin susceptibility in both the normal and superconducting state of 
Ba$_{0.67}$K$_{0.33}$Fe$_2$As$_{2}$, different from optimally electron-doped BaFe$_{1.88}$Co$_{0.12}$As$_2$
where the anisotropy is believed to emerge only with the opening of the superconducting gap \cite{steffens}.
Furthermore, our data reveal that large differences in the superconductivity-induced 
spin gaps at ${\bf Q}=(0.5,0.5,0)$ and
$(0.5,0.5,1)$ \cite{clzhang} arise from the differences in spin gaps of $c$-axis polarized spin excitations.  These results are similar to the previous work on 
electron-doped BaFe$_{1.9}$Ni$_{0.1}$As$_{2}$ \cite{lipscombe10} and BaFe$_{1.88}$Co$_{0.12}$As$_2$ \cite{steffens}, suggesting that the influence of a strong spin anisotropy in undoped parent compound 
BaFe$_2$As$_2$ \cite{qureshi12} extends to both optimally electron and hole-doped iron pnictide superconductors. 
For comparison, we note that spin excitations in superconducting iron chalcogenides are different, having
slightly anisotropic resonance with isotropic spin excitations below the resonance \cite{boothroyd,prokes}.

In Ref. \cite{steffens}, it was suggested that the observed spin anisotropy 
in BaFe$_{1.88}$Co$_{0.12}$As$_2$
can be understood as a $c$-axis polarized resonance whose intensity strongly varies with the $c$-axis wave vector.  This is not the case in Ba$_{0.67}$K$_{0.33}$Fe$_2$As$_{2}$ since we find much weaker $c$-axis modulation of
the magnetic intensity \cite{clzhang}.  Therefore, the spin anisotropy seen in optimally electron and hole-doped superconductors is a consequence of these materials being close to the AF ordered parent compound BaFe$_2$As$_2$, where spin-orbit coupling is expected to be strong \cite{kruger,lee,lv}, and is not fundamental to superconductivity of these materials.  To understand how spin anisotropy 
in optimally hole and electron-doped iron pnictide 
superconductors might be coupled to superconductivity via spin-orbit interaction, we note
that hole and electron-doped iron pnictides are multiband superconductors with different superconducting gaps
for different orbitals.  If $c$-axis and in-plane spin excitations 
arise from quasiparticle excitations of different orbitals between hole and 
electron Fermi pockets \cite{jhzhang}, the large differences in superconducting gaps for Fermi surfaces of
different orbital characters might induce the observed large spin anisotropy.  
  
We are grateful to W. C. Lv for helpful discussions and H.F. Li, K. Schmalzl, and W. Schmidt for their assistance in the neutron scattering experiment. The work at UTK is supported by the US DOE BES No. DE-FG02-05ER46202. 
C.L.Z and T.E are partially supported by the US DOE BES through the EPSCoR grant, DE-FG02-08ER46528. 
Work at IOP is supported by the MOST of China 973 programs (2012CB821400).

\end{document}